\documentclass[fleqn,twoside]{article}

\usepackage{espcrc2}

\usepackage{amsmath,amstext,amsfonts,amsbsy,amssymb,amscd,bbm,epsfig}
\usepackage{graphicx}
\usepackage{subfigure}

\newcommand{\ba}{\begin{array}}
\newcommand{\ea}{\end{array}}

\newcommand{\req}[1]{Eq.~(\ref{#1})}

\newcommand{\rep}[1]{\cite{#1}}
\newcommand{\refig}[1]{Fig.~\ref{#1}}
\newcommand{\ret}[1]{Table~\ref{#1}}

\newcommand{\dif}{{\rm d}}

\newcommand{\Dslash}{\relax{\kern+.25em / \kern-.70em D}}

\newcommand{\Real}{\relax{\mathsf{\Gamma\kern-.35em R}}}
\newcommand{\Int}{\relax{\mathsf{Z\kern-.40em Z}}}

\newcommand{\MSbar}{\overline{\rm MS}}

\newcommand{\gbar}{\kern1pt\overline{\kern-1pt g\kern-0pt}\kern1pt}
\newcommand{\mbar}{\kern2pt\overline{\kern-1pt m\kern-1pt}\kern1pt}
\newcommand{\obar}[1]{\kern3pt\overline{\kern-2pt #1\kern-0pt}\kern1pt}

\newcommand{\lmax}{L_{\rm max}}

\newcommand{\Oa}{\mbox{O}(a)}

\newcommand{\cC}{{\cal C}}

\newcommand{\cK}{{\cal K}}

\newcommand{\cO}{{\cal O}}
\newcommand{\cP}{{\cal P}}

\newcommand{\cS}{{\cal S}}

\newcommand{\vx}{\mathbf{x}}
\newcommand{\vy}{\mathbf{y}}

\usepackage{subfigure}

\usepackage{graphicx}

\newcommand{\eqindent}{\relax{\kern-20pt}}
    \title{
    \vspace{-25mm}
    \rightline{\small ROM2F/2003/20~~~~~}
    \rightline{\small DESY 03-129~~~~~}
    \rightline{\small MS-TP-03-10~~~~~}
    \rightline{\small IFT UAM-CSIC/03-34~~~~~}
    $B_K$ from twisted mass QCD
    \vspace*{-2mm}
    \leftline{\small ALPHA Collaboration~~~~}}
    \author{P. Dimopoulos\address[ToV]{INFN, Sezione di Roma II,
    Dipartimento di Fisica, Universit\`a di Roma ``Tor Vergata'',
    Via della Ricerca scientifica 1, I-00133, Italy}
    \thanks{Based on a poster presented by P. Dimopoulos at the  LATTICE 2003 Conference  (Tsukuba, Japan).}\setcounter{footnote}{0},
      J.~Heitger \address{Westf\"alische Wilhelms-Universit\"at M\"unster,
        Institut f\"ur Theoretische Physik, Wilhelm-Klemm-Strasse 9,
        D-48149 M\"unster, Germany},
      C.~Pena\address{DESY, Theory Group, Notkestrasse 85, D-22607 Hamburg, Germany},
      S.~Sint \address{Departamento de F\'{\i}sica Te\'orica C-XI and
        Instituto de F\'{\i}sica Te\'orica C-XVI, Universidad
        Aut\'onoma de Madrid, Cantoblanco E-28049 Madrid, Spain} and
      A.~Vladikas\addressmark[ToV]}

\begin{document}

\begin{abstract}
We present some preliminary results for $B_K$ at $\beta=6.0$, using the twisted
mass QCD formalism for the computation of bare matrix elements of the
$\Delta S=2$ operator. The main advantage of the method is that mixing
with other $d=6$ operators under renormalisation is avoided. Moreover
the operator renormalisation is performed in the Schr\"odinger
functional (SF) framework, using earlier results of our collaboration
for the corresponding step scaling function.
\end{abstract}

\maketitle

\section{$B_K$ and tmQCD}

Let us start by recalling in brief the tmQCD framework
for the computation of $B_K$~\rep{BKtmQCD}. We start
from a continuum fermion action of the form
\begin{align}\eqindent
  S_{\rm F} = \int\dif^4 x\,\Big\{&
    \bar\psi(x)\left[\Dslash+m_l+i\mu_l\gamma_5\tau_3\right]\psi(x) \nonumber  \    \\ 
    &+\bar s(x)\left[\Dslash+m_s\right]s(x)\Big\} \ ,
\end{align}
where $\psi=(u,d)^T$ is a doublet of degenerate light quarks, and
$\tau_3$ acts on isospin space. The axial transformation
$\psi \to e^{i\alpha\gamma_5\tau_3/2}\psi~,~~
\bar\psi \to \bar\psi\, e^{i\alpha\gamma_5\tau_3/2}$
leaves the form of the action invariant, and induces a rotational
transformation of the mass parameters.
In particular, by choosing the rotation angle such that $\tan(\alpha)=\mu_l/m_l$
the standard QCD action is recovered. The quantity
$M_l=\sqrt{m_l^2+\mu_l^2}$ is left invariant, and can
be identified with the physical light quark mass.

The above chiral rotation, on the other hand, can be seen
as a change of fermion variables which induces a mapping between composite
operators. If we denote with a prime the quantities in standard QCD variables
then one has, in particular,
\begin{gather}
  A'_\mu = \cos\left(\frac{\alpha}{2}\right)A_\mu -
  i\sin\left(\frac{\alpha}{2}\right)V_\mu \ , \\
  \relax{\kern+2pt}V'_\mu = \cos\left(\frac{\alpha}{2}\right)V_\mu -
  i\sin\left(\frac{\alpha}{2}\right)A_\mu \ , \\
  \label{oper_mapping}
  \relax{\kern-8mm}O'^{\Delta S=2}_{\scriptscriptstyle\rm VV+AA} =
  \cos(\alpha)O^{\Delta S=2}_{\scriptscriptstyle\rm VV+AA} 
  - i\sin(\alpha)O^{\Delta S=2}_{\scriptscriptstyle\rm VA+AV} \ ,
\end{gather}
where all the operators have an $s-d$ flavour structure.
These tree-level properties are immediately extended to the renormalised
quantum theory, provided renormalised quark masses and composite operators
are used in the above relations. Then~\req{oper_mapping} implies that,
for the special case $\alpha=\pi/2$, the physical matrix
element of $O^{\Delta S=2}$ entering $B_K$ is given in the twisted theory
by a matrix element of the VA+AV part of the operator, which is protected
from mixing under renormalisation by $\cC\cP\cS$ symmetry and therefore
renormalises multiplicatively~\rep{bernard}.

To compute $B_K$ we regularise the theory using a Schr\"odinger
Functional (SF) action with Wilson fermions. The action is
nonperturbatively $\Oa$ improved in the bulk of the SF cylinder, and
one-loop $\Oa$ improved at the time boundaries. The bare value of $B_K$ is
extracted from a fit to the central plateau of the ratio of correlation functions
\begin{gather}\eqindent
  \label{ratio_bk}
  R_{B_K}(x_0) = \frac
  {-i\langle\cO'O^{\Delta S=2}_{\rm VA+AV}\cO\rangle}
  {\frac{8}{3}\langle\cO'\cK(x)\rangle\langle\cK(x)\cO\rangle} \ ,
\end{gather}
where $\cK=\frac{1}{\sqrt{2}}\left[Z_A A_0-iZ_V V_0\right]$
is the combination of renormalised operators which represents the
physical axial current in twisted variables, and
$\cO= L^{-3}\sum_{\vx,\vy}\bar\zeta_d(\vx)\gamma_5\zeta_s(\vy)$
is a pseudoscalar SF boundary source\footnote{In~\req{ratio_bk},
  in standard SF fashion, the prime denotes the $x_0=T$ boundary.}.

\begin{table}[t]
 \centering
 \begin{tabular}{cccc}
 \hline\\[-1.8ex]
 Set &  $aM_{\rm PS}$& $aM'_{\rm PS}$ & $B_K$ \\
 \hline\\[-1.8ex]
 I   & $0.3901(15)$ & $0.3892(16)$ & $1.022(11)$ \\
 II  & $0.3561(20)$ & $0.3546(20)$ & $0.999(24)$ \\
 III & $0.3298(19)$ & $0.3284(19)$ & $0.984(23)$ \\
 IV  & $0.3175(23)$ & $0.3040(24)$ & $1.003(27)$ \\
 \hline\\[-1.8ex]
 && Extrap. & 0.959(64) \\
 \hline\\
 \end{tabular}
 \caption{Values of the bare parameter $B_K$ and pseudoscalar masses
 for the four sets of quark masses and extrapolation of  $B_K$ to
 the physical kaon mass. (Sets correspond to  245, 150, 150, 102 gauge configurations respectively.)}
 \label{tab:bk}
\vspace{-8mm}
\end{table}

\section{Results}

Here we present results at $\beta=6.0$, all of them on a
$16^3 \times 48$ lattice, corresponding to an approximate physical volume
of $(1.5~{\rm fm})^3 \times 4.5~{\rm fm}$. All the simulations have been
carried out on the APEmille machines at DESY-Zeuthen. Four sets of quark
mass values have been considered. Three of them correspond to degenerate
$s-d$ quark masses~\footnote{The degeneracy holds for renormalised
  quark masses and up to $\cO(a^2)$ cutoff effects.}, and the fourth
one introduces a small ($\approx 15\%$) $s-d$ isospin breaking.

The lowest pseudoscalar meson mass can be extracted either from an $s-d$
two-point correlation function with the appropriate quantum numbers or,
alternatively, from an $s-d'$ correlation function, where $d'$ is an
untwisted down quark with exactly the same mass as $s$.
In~\ret{tab:bk} (2nd and 3rd columns) we show the values $M_{\rm PS}$
and $M'_{\rm PS}$ for the corresponding pseudoscalar masses.
If the twisted $d$ quark is degenerate in mass with the $s$ up to
$\cO(a^2)$ cutoff effects, then also the two meson masses should
differ only by $\cO(a^2)$ effects. This is clearly seen in the data,
which show no noticeable difference between $M_{\rm PS}$
and $M'_{\rm PS}$ in the first three cases, whereas in the fourth case
isospin breaking manifests itself in the non--equality of the masses.

\begin{figure}[t]
\centering
\includegraphics[scale=0.28,angle=-90]{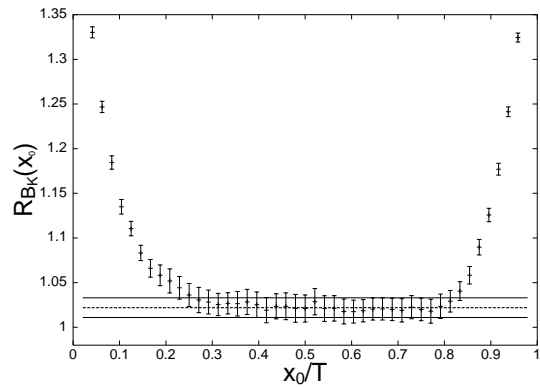}\vspace{-5mm}
\caption{$R_{B_K}(x_0)$ and the fit to the plateau for $B_K$ for the
  data set I of~\ret{tab:bk}.}
\label{figks1}
\vspace{-9mm}
\end{figure}

In~\ret{tab:bk} (4th column) we present our results for the bare value of $B_K$ for
each parameter set, as well as the result obtained by extrapolation to the
physical kaon mass, which we take to be $aM_K=0.2336$. In all cases the
fitting interval is $x_0/a\in[16,32]$.
As an example in Fig.~\ref{figks1} we depict  the ratio $R_{B_K}$ and
the fit to the plateau which gives the $B_K$ for the data set I.

\begin{figure}[t]
\centering
\includegraphics[scale=0.30,angle=-90]{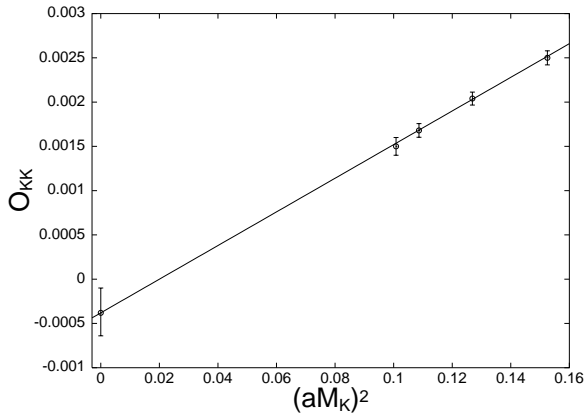}\vspace{-5mm}
\caption{Extrapolation to the chiral limit of the $K^0-\bar{K^0}$ matrix element.}  
\label{ME}
\vspace{-8mm}
\end{figure}

The data also allow to examine the chiral behaviour of the
$\langle\bar K^0 \vert O^{\Delta S = 2} \vert K^0 \rangle$ matrix
element, which can be directly computed from the ratio
\begin{gather}\eqindent
  R_{O_{KK}}(x_0) = \frac
  {-i\langle\cO'O^{\Delta S=2}_{\rm VA+AV}\cO\rangle}
  {\langle\cO'\cO\rangle} \ ,
\end{gather}
and then studied as a function of $(a M_K)^2$. The result is shown in
Fig.~\ref{ME}. The operator $O_{\scriptscriptstyle\rm VA+AV}^{\Delta S = 2}$ has
been renormalised in the first of the SF  schemes
discussed in the next section, at the hadronic scale $\mu = 1/(2 L_{{\rm max}})$.
The intercept of the linear extrapolation is $C = - 0.00038(27)$,
i.e. compatible with zero within 1.5 standard deviations. We interpret this
as a signal of correct chiral behaviour, in view of the fact that
the prediction of a vanishing matrix element in the chiral limit
is only true for this quantity in the continuum. Here we are working with
an unimproved matrix element computed at fixed lattice spacing, and
the behaviour in~\refig{ME} is still affected by mild $\cO(a)$ effects.
Moreover, as in previous studies with Wilson fermions, the data lie in a
region where LO chiral perturbation theory predictions have to be taken
with care.

\section{Renormalisation}

For the nonperturbative renormalisation of $O^{\Delta S=2}_{\rm VA+AV}$ we
use SF techniques, as described in~\rep{ssf_proc}. Nine different SF
renormalisation schemes are available, each one of them supplying a
formally (up to statistical correlations) independent
determination of the renormalisation group invariant parameter
$B_K^{\rm RGI}$. Having only $\beta=6.0$ results
and thus no continuum extrapolation yet,
an estimate  of the latter is obtained
by multiplying the bare value of $B_K$ times the renormalisation constant
at $\beta=6.0$ and renormalisation scale $1/2\lmax$, after which the result
is multiplied by the continuum quantity $B_K^{\rm RGI}/\bar B_K^{\rm
  SF}(1/(2\lmax))$, obtained using the corresponding nonperturbative step
scaling function and NLO perturbation theory (see~\rep{ssf_proc} for
details). The renormalisation constant and the ratio are scheme-dependent
quantities, whereas the final result is not. In~\ret{tab:renorm} we supply
the resulting estimate for $B_K^{\rm RGI}$ and
$\bar B_K^{\MSbar}(2~{\rm GeV})$ in each of the nine schemes.

The combination of the nine estimates of $B_K^{\rm RGI}$ will
ultimately yield a good control over systematic effects and a
reduction of the uncertainty of the final result. It has to be
stressed, though, that the results in~\ret{tab:renorm}, while being
compatible within errors, involve different cutoff effects, and hence
a reliable answer can be obtained only after extrapolation to the
continuum limit (simulations at higher values of $\beta$ will follow
soon). Note, however, that the $\beta=6.0$ result is already close to the
continuum limit values found in the literature, which suggests that cutoff
results in our framework might turn out to be remarkably small.


\begin{table}[t]
 \centering
 \begin{tabular}{c@{\hspace{10mm}}c@{\hspace{5mm}}c}
 \hline\\[-1.8ex]
 Scheme & $B_K^{\rm RGI}$ & $\bar B_K^{\MSbar}(2~{\rm GeV})$ \\
 \hline\\[-1.8ex]
 1  & 0.88(7)& 0.64(5)\\
 2a & 0.94(8)& 0.68(6)\\
 2b & 0.92(8)& 0.67(6)\\
 3a & 0.89(7)& 0.64(5)\\
 3b & 0.87(7)& 0.63(5)\\
 4a & 0.94(7)& 0.68(5)\\
 4b & 0.93(8)& 0.67(6)\\
 5a & 0.94(8)& 0.68(6)\\
 5b & 0.93(8)& 0.67(6)\\
 \hline\\
 \end{tabular}
 \caption{Values for $B_K^{\rm RGI}$ and
   $\bar B_K^{\MSbar}(2~ {\rm GeV})$ obtained after renormalisation at $\beta=6.0$.}
 \label{tab:renorm}
\vspace{-9mm}
\end{table}

\vspace*{-0.35cm}
\section*{ACKNOWLEDGEMENTS}
\vspace*{-0.25cm}
We wish to thank D. Becirevic, L. Giusti, M. Guagnelli  and R. Sommer for useful discussions. 
This work was supported in part by the European Community's Human
Potential Programme under contract HPRN-CT-2000-00145, Hadrons/Lattice
QCD.




\end{document}